\documentclass[article]{elsarticle}

\usepackage{lineno,hyperref}
\usepackage{mathptmx}
\usepackage{amsmath}
\usepackage{amssymb}
\usepackage{sectsty}
\usepackage{graphicx} 
\usepackage{color}
\usepackage{colortbl}
\usepackage[usenames,dvipsnames]{xcolor}

\journal{Preprint}
\begin{document}

\begin{frontmatter}

%\title{A glimpse of ocean of abundant discoveries : Two-way cross
%  family analysis of in-silico ranked $2^{nd}$ order unexplored,
%  ETC-1922159 affected, synergistic combinations in CRC cells} 

\title{Machine learning discoveries of ASCL2-X synergy in ETC-1922159 treated
  colorectal cancer cells}

\tnotetext[mytitlenote]{ML dicoveries of ASCL2-X synergy in ETC-1922159 treated
  CRC cells}

\begin{abstract}
Achaete-scute complex homolog 2 (ASCL2) codes a part of the basic helix-loop-helix (BHLH) transcription factor family. WNTs have been found to directly affect the stemness of the tumor cells via regulation of ASCL2. Switching off the ASCL2 literally blocks the stemness process of the tumor cells and vice versa. In colorectal cancer (CRC)
cells treated with ETC-1922159, ASCL2 was found to be down regulated
along with other genes. A recently developed search engine ranked
combinations of ASCL2-X (X, a 
particular gene/protein) at $2^{nd}$ order level after drug
administration. Some rankings confirm the already tested combinations, while others point to those that are untested/unexplored. These rankings reveal which ASCL2-X combinations might
be working synergistically in CRC. In this research work,
I cover combinations of ASCL2 with WNT, transforming growth factor $\beta$ (TGF$\beta$), interleukin (IL), leucine rich repeat containing G protein-coupled receptor (LGR), NOTCH, solute carrier family (SLC), SRY-box transcription factor (SOX), small nucleolar RNA host gene (SNHG), KIAA, F-box protein (FBXO), family with sequence similarity (FAM), B cell CLL/lymphoma (BCL), autophagy related (ATG) and Rho GTPase activating protein (ARHGAP) family. 
\end{abstract} 

%% Group authors per affiliation:
\author{shriprakash sinha}
%shriprakash sinha
\address{Independent Researcher; Orcid ID :  orcid.org/0000-0001-7027-5788}
\address{104-Madhurisha Heights Phase 1, Risali, Bhilai-490006,
India}
%Independent Researcher; Orcid ID :  orcid.org/0000-0001-7027-5788
\fntext[myfootnote]{Aspects of unpublished work were presented in a poster session
  at (1) the recently concluded first ever Wnt Gordon Conference, from 6-11 August 2017,
  held in Stowe, VT 05672, USA.}
\ead{sinha.shriprakash@yandex.com}
%sinha.shriprakash@yandex.com

%% or include affiliations in footnotes:
%\author[mymainaddress,mysecondaryaddress]{Elsevier Inc}
%\ead[url]{www.elsevier.com}

%\author[mysecondaryaddress]{Global Customer Service\corref{mycorrespondingauthor}}
%\cortext[mycorrespondingauthor]{Corresponding author}
%\ead{support@elsevier.com}

%\address[mymainaddress]{1600 John F Kennedy Boulevard, Philadelphia}
%\address[mysecondaryaddress]{360 Park Avenue South, New York}

\end{frontmatter}

%\linenumbers
%\tableofcontents

\section{Introduction}
\subsection{Stemness and intestine}
Stemness is a property by which the cells proliferate, regenerate, sustain, propogation. It is the property by which the cells are known to maintain distinctive aspects over their lineage. Stemness is a property that is observed in both normal and cancer stem cells. Cancer stem cells show high self renewal, redundancy in self renewed pathways leading to pathological conditions, genomic instability due to altered DNA repair factors and epithelial mesenchymal transition as a measure of plasticity. \citet{Aponte:2017stemness} review the stemness in cancer cells in great detail. \par
The human intestine is broadly divided into small and large intestine for identification and study. The epithelium of the small intestine is composed of an array of villus and crypts (mountains and valleys for the uninitiated). The crypts are also termed as crypt of Lieberkuhn after his study of the intestine in this PhD thesis \citet{Joa:1780dissertatio}. The large intestine or the colon, however does not contain the villi. The multiple types of cells available in the intestinal glands at the villus are : enterocytes for absorbing water and electrolytes, goblet cells for secreting mucus, enteroendocrine cells for secreting hormones, cup cells, tuft cells and at the crypt of the gland are : paneth cells for secreting anti-microbial peptides for maintenance of gastrointestinal tract and stem cells. These cells are not all present in the colon. The description of the functioning of the cells along the longitudinal axis from the histological perspective of the intestine has been explained elaborately in \citet{Potten:2002intestinal}, \citet{Barker:2012identifying}, \citet{Clevers:2013intestinal} and \citet{Noah:2011intestinal}. Briefly, the stem cells help in the production of the transit population of the progenitor cells that differentiate as they move towards the villus. Once the cells are committed for differentiation, they no longer have the property of stem cells.\par
The Wnt pathway is implicated in the control of stemness and cell fate in the intestine. A family of the WNT proteins, WNT10B, might be playing a crucial role in stemness. This is further confirmed by wet lab experiments in \citet{Reddy:2016},
which show BVES deletion results in amplified stem cell activity and
Wnt signaling after radiation. WNT10B has also been implicated in colorectal cancer \citet{Yoshikawa:2007}. \par
\subsection{Achaete-scute like 2 (ASCL2)}
ASCL2 belongs to a family of ASCL genes that contain basic and helix-loop-helix domains in a conserved family of transcriptional regulators. ASCL2 is known to be genomically imprinted at the cluster location of chromosome 11p15.5. Genomic imprinting is the reversible epigenetic silencing of the parental specific inherited genes \citet{Schwienbacher:2000gain}. ASCL2 has been found to be maternally expressed and \citet{Miyamoto:2002human} show the expression patterns of the human ASCL2 in the fetus at a stage between first and second trimesters and in the placental tissues. In addition, they demonstrate that the ASCL2 escapes genomic imprinting during the mentioned stages. In colorectal cancer, loss of this genomic imprinting (i.e silencing) has been found to cause overexpression of ASCL2 \citet{Cui:1998loss} \& \citet{Stange:2010expression}. In intestinal neoplasia, ASCL2 is found to be a target of Wnt signaling \citet{Jubb:2006achaete}. \citet{Jubb:2006achaete} report that mutiple strategies have been chosen to repress the Wnt signaling pathway and found that expression of ASCL2 was repressed. Also, ASCL2 was found not to be imprinted in neoplasia and knockdown of ASCL2 led to cell cycle (G2/M) arrest which affects the expression of Survivin and cdc25b, which are both found to be highly expressed in colorectal cancer. \citet{Jubb:2006achaete} further show that ASCL2 expression controls G2/M checkpoint and promotes proliferation by initiating the transcriptional regulation of Survivin and cdc25b which are targets of Wnt signaling. This is further proported by ASCL2 knockdown experiments in \citet{Zhu:2012ascl2}. \par
\subsection{Combinatorial search problem and a possible solution}
In a recently published work \citet{Sinha:2024machine}, a
frame work of a search engine was developed which can rank combinations
of factors (genes/proteins) in a signaling pathway. Readers are requested to go
through the adaptation of the above mentioned work for gaining deeper insight into the working of the
pipeline and its use of published data set generated after
administration of ETC-1922159, \citet{Sinha:2017b}. The work uses SVM
package by \citet{Joachims:2006} in
\url{https://www.cs.cornell.edu/people/tj/svm_light/svm_rank.html}. I
use the adaptation to rank 2$^{nd}$ order gene combinations. \par
\section{Results \& Discussion}
\subsection{ASCL2 related synergies}
\subsection{ASCL2 - WNT10B}
ASCL2 has been found to play a major role in stemness in colon crypts and is
implicated in colon cancer \citet{Zhu:2012}. Switching off the ASCL2
leads to a literal blockage of the stemness process and vice versa. At
the downstream level, ASCL2 is regulated by TCF4/$\beta$-catenin via
non-coding RNA target named WiNTRLINC1
\citet{Giakountis:2016}. Activation of ASCL2 leads to feedforward
transcription of the non-coding RNA and thus a loop is formed which
helps in the stemness and is highly effective in colon cancer. At the
upstream level, ASCL2 is known act as a WNT/RSPONDIN switch that
controls the stemness \citet{Schuijers:2015}. It has been shown that
removal of RSPO1 lead to decrease in the Wnt signaling due to removal
of the FZD receptors that led to reduced expression of ASCL2. Also,
low levels of LGR5 were observed due to this phenomena. The opposite
happened by increasing the RSPO1 levels. After the drug treatment, it
was found that ASCL2 was highly suppressed pointing to the inhibition
of stemness in the colorectal cancer cells. Also,
\citet{Schuijers:2015} show that by genetically disrupting PORCN or
inducing a PORCN inhibitor (like IWP-2), there is loss of stem cell
markers like LGR5 and RNF43, which lead to disappearance of stem cells
and moribund state of mice. A similar affect can be found with
ETC-1922159, where there is suppression of RNF43 and LGR5 that lead to
inhibition of the Wnt pathway and thus the ASCL2 regulation. These wet
lab evidences are confirmed in the relatively low ranking of the
combination ASCL2-RNF43 via the inhibition of PORCN-WNT that leads to
blocking of the stemness that is induced by ASCL2. Since ASCL2 is
directly mediated by the WNT proteins, the recorded ASCL2-WNT10B combination showed
low priority ranking of 497, 321 and 488 for laplace, linear and rbf kernels,
respectively, thus indicating a possible connection between WNT10B and ASCL2
activation (data not shown in tabular format). \par
\subsubsection{ASCL2 - TGFB}
\citet{Zhang:2023ascl2} showed that overexpression of ASCL2 increased TGFB levels thus stimulating local cancer associated fibroblasts activation. This induced an immune-excluded microenvironment. In colorectal cancer cells treated with ETC-1922159, TGFB family and ASCL2, were found to be down regulated and recorded independently. I was able to rank 2$^{nd}$ order combination of
TGFB family and ASCL2, that were down regulated. \par
Table \ref{tab:ASCL2-TGFB-interaction} shows rankings of these
combinations. Followed by this is the unexplored
combinatorial hypotheses in table \ref{tab:ASCL2-TGFB-hypotheses}
generated from analysis of the ranks in table
\ref{tab:ASCL2-TGFB-interaction}. The table
\ref{tab:ASCL2-TGFB-interaction} shows rankings of TGFB family w.r.t
ASCL2. TGFBR3 - ASCL2 shows low ranking of 819 (laplce), 1072 (linear) and 1170 (rbf). These rankings point to the synergy existing between the two components, which have been down regulated after the drug treatment. Further, TGFB1 and TGFBRAP1 showed high ranking and might not be synergistically working with ASCL2, before treatment. \par
\begin{table}[ht!]
\begin{center}
\resizebox{\columnwidth}{!}{
\begin{tabular}{@{}lclc@{}}
\multicolumn{4}{c}{\textsc{Ranking TGFB family VS ASCL2}}\\
\hline
\multicolumn{4}{c}{\textsc{Ranking of TGFB family w.r.t ASCL2}} \\
                 & laplace  & linear & rbf \\
\hline
TGFBR3 - ASCL2 & 819 &  1072  &  1170  \\
TGFB1 - ASCL2 & 2210 &  2378  &  1754 \\
TGFBRAP1 - ASCL2 & 2467 &  1978  &  2738 \\
\hline
\end{tabular}
} % end of resizebox
\end{center}
\caption{$2^{nd}$ order interaction ranking between ASCL2 VS TGFB family
members.}
\label{tab:ASCL2-TGFB-interaction}
\end{table}
One can also interpret the results of the table
\ref{tab:ASCL2-TGFB-interaction} graphically, with the following
influences - $\bullet$ TGFB family w.r.t ASCL2 with ASCL2 $->$ TGFB-R3. \par
\begin{table}[h!]
\begin{center}
\resizebox{\columnwidth}{!}{
\begin{tabular}{@{}ll@{}}
\multicolumn{2}{c}{\textsc{Unexplored combinatorial hypotheses}}\\
\hline
TGFB family w.r.t ASCL2 & \\
TGFB-R3 & ASCL2\\
\hline
\end{tabular}
} % end of resizebox
\end{center}
\caption{$2^{nd}$ order combinatorial hypotheses between ASCL2
  and TGFB family members.}
\label{tab:ASCL2-TGFB-hypotheses}
\end{table}
%
%\clearpage
%
\subsubsection{ASCL2 - IL}
Evaluation of ASCL2 bound promoters by \citet{Murata:2020ascl2} in regenerating cells, showed that interleukin-11 receptor gene IL-11RA1 as a transcriptional target, and respective organoid cultures demonstrated IL-11 activity in intestinal stem cell regeneration. In colorectal cancer cells treated with ETC-1922159, IL family and ASCL2, were found to be down regulated and
recorded independently. I was able to rank 2$^{nd}$ order combination of
IL family and ASCL2, that were down regulated. \par
Table \ref{tab:ASCL2-IL-interaction} shows rankings of these
combinations. Followed by this is the unexplored
combinatorial hypotheses in table \ref{tab:ASCL2-IL-hypotheses}
generated from analysis of the ranks in table
\ref{tab:ASCL2-IL-interaction}. The table
\ref{tab:ASCL2-IL-interaction} shows rankings of IL family w.r.t
ASCL2. IL17D - ASCL2 shows low ranking of 279 (laplace), 805 (linear) and 1318 (rbf). IL17RB - ASCL2 shows low ranking of 519 (laplace), 609 (linear) and 1026 (rbf). IL33 - ASCL2 shows low ranking of 754 (laplace), 417 (linear) and 119 (rbf). IL1RL2 - ASCL2 shows low ranking of 1342 (laplace) and 1563 (rbf). ILF3 - ASCL2 shows low ranking of 1397 (laplace) and 1421 (linear). These rankings point to the synergy existing between the two components, which have been down regulated after the drug treatment. Further, ILF3-AS1, IL17RD and ILF2 showed high ranking and might not be synergistically working with ASCL2, before treatment. \par
\begin{table}[ht!]
\begin{center}
\resizebox{\columnwidth}{!}{
\begin{tabular}{@{}lclc@{}}
\multicolumn{4}{c}{\textsc{Ranking IL family VS ASCL2}}\\
\hline
\multicolumn{4}{c}{\textsc{Ranking of IL family w.r.t ASCL2}} \\
                 & laplace  & linear & rbf \\
\hline
IL17D - ASCL2 & 279 &  805  &  1318  \\
IL17RB - ASCL2 & 519 &  609  &  1026 \\
IL33 - ASCL2 & 754 &  417  &  119 \\
IL1RL2 - ASCL2 & 1342 &  1860  &  1563 \\
ILF3 - ASCL2 & 1397 &  1421  &  1878 \\
ILF3-AS1 - ASCL2 & 1720 &  764  &  2113 \\
IL17RD - ASCL2 & 2444 &  2331  &  2263 \\
ILF2 - ASCL2 & 2683 &  2329  &  2154 \\
\hline
\end{tabular}
} % end of resizebox
\end{center}
\caption{$2^{nd}$ order interaction ranking between ASCL2 VS IL family
members.}
\label{tab:ASCL2-IL-interaction}
\end{table}
One can also interpret the results of the table
\ref{tab:ASCL2-IL-interaction} graphically, with the following
influences - $\bullet$ IL family w.r.t ASCL2 with ASCL2 $->$ IL-17DR3/17RB/33/1RL2/F3. \par
\begin{table}[h!]
\begin{center}
\resizebox{\columnwidth}{!}{
\begin{tabular}{@{}ll@{}}
\multicolumn{2}{c}{\textsc{Unexplored combinatorial hypotheses}}\\
\hline
IL family w.r.t ASCL2 & \\
IL-17DR3/17RB/33/1RL2/F3 & ASCL2\\
\hline
\end{tabular}
} % end of resizebox
\end{center}
\caption{$2^{nd}$ order combinatorial hypotheses between ASCL2
  and IL family members.}
\label{tab:ASCL2-IL-hypotheses}
\end{table}
%
%\clearpage
%
\subsubsection{ASCL2 - LGR}
In gastric cancer, \citet{Kwon:2013aberrant} show that ASCL2 upregulates LGR5 expression at transcriptional level. In colorectal cancer cells treated with ETC-1922159, LGR family and ASCL2, were found to be down regulated and recorded independently. I was able to rank 2$^{nd}$ order combination of LGR family and ASCL2, that were down regulated. \par
Table \ref{tab:ASCL2-LGR-interaction} shows rankings of these
combinations. Followed by this is the unexplored
combinatorial hypotheses in table \ref{tab:ASCL2-LGR-hypotheses}
generated from analysis of the ranks in table
\ref{tab:ASCL2-LGR-interaction}. The table
\ref{tab:ASCL2-LGR-interaction} shows rankings of LGR family w.r.t
ASCL2. LGR5 - ASCL2 shows low ranking of 70 (laplace), 85 (linear) and 213 (rbf). LGR6 - ASCL2 shows low ranking of 304 (laplace), 463 (linear) and 964 (rbf). These rankings point to the synergy existing between the two components, which have been down regulated after the drug treatment. \par
\begin{table}[ht!]
\begin{center}
\resizebox{\columnwidth}{!}{
\begin{tabular}{@{}lclc@{}}
\multicolumn{4}{c}{\textsc{Ranking LGR family VS ASCL2}}\\
\hline
\multicolumn{4}{c}{\textsc{Ranking of LGR family w.r.t ASCL2}} \\
                 & laplace  & linear & rbf \\
\hline
LGR5 - ASCL2 & 70 & 85  &  213  \\
LGR6 - ASCL2 & 304 & 463  &  964 \\
\hline
\end{tabular}
} % end of resizebox
\end{center}
\caption{$2^{nd}$ order interaction ranking between ASCL2 VS LGR family
members.}
\label{tab:ASCL2-LGR-interaction}
\end{table}
One can also interpret the results of the table
\ref{tab:ASCL2-LGR-interaction} graphically, with the following
influences - $\bullet$ LGR family w.r.t ASCL2 with ASCL2 $->$ LGR-5/6. \par
\begin{table}[h!]
\begin{center}
\resizebox{\columnwidth}{!}{
\begin{tabular}{@{}ll@{}}
\multicolumn{2}{c}{\textsc{Unexplored combinatorial hypotheses}}\\
\hline
LGR family w.r.t ASCL2 & \\
LGR-5/6 & ASCL2\\
\hline
\end{tabular}
} % end of resizebox
\end{center}
\caption{$2^{nd}$ order combinatorial hypotheses between ASCL2
  and LGR family members.}
\label{tab:ASCL2-LGR-hypotheses}
\end{table}
%
%\clearpage
%
\subsubsection{ASCL2 - NOTCH}
In spinous cells, based upon the ability of NOTCH/RBP-J to bind to the promoter region of ASCL2, \citet{Moriyama:2008multiple} speculated that activation of NOTCH signaling promotes the expression of both the HES1 and ASCL2 genes. However, subsequent HES1 binding to the ASCL2 promoter interferes with NOTCH/RBP-J-induced ASCL2 expression, thus resulting in inhibition of ASCL2. In colorectal cancer cells treated with ETC-1922159, NOTCH family and ASCL2, were found to be down regulated and recorded independently. I was able to rank 2$^{nd}$ order combination of NOTCH family and ASCL2, that were down regulated. \par
Table \ref{tab:ASCL2-NOTCH-interaction} shows rankings of these
combinations. Followed by this is the unexplored
combinatorial hypotheses in table \ref{tab:ASCL2-NOTCH-hypotheses}
generated from analysis of the ranks in table
\ref{tab:ASCL2-NOTCH-interaction}. The table
\ref{tab:ASCL2-NOTCH-interaction} shows rankings of NOTCH family w.r.t
ASCL2. NOTCH1 - ASCL2 shows low ranking of 748 (laplace) and 898 (rbf). NOTCH4 - ASCL2 shows low ranking of 1021 (laplace), 378 (linear) and 873 (rbf). These rankings point to the synergy existing between the two components, which have been down regulated after the drug treatment. \par
\begin{table}[ht!]
\begin{center}
\resizebox{\columnwidth}{!}{
\begin{tabular}{@{}lclc@{}}
\multicolumn{4}{c}{\textsc{Ranking NOTCH family VS ASCL2}}\\
\hline
\multicolumn{4}{c}{\textsc{Ranking of NOTCH family w.r.t ASCL2}} \\
                 & laplace  & linear & rbf \\
\hline
NOTCH1 - ASCL2 & 748 & 1745 & 898  \\
NOTCH4 - ASCL2 & 1021 & 378 &  873 \\
\hline
\end{tabular}
} % end of resizebox
\end{center}
\caption{$2^{nd}$ order interaction ranking between ASCL2 VS NOTCH family
members.}
\label{tab:ASCL2-NOTCH-interaction}
\end{table}
One can also interpret the results of the table
\ref{tab:ASCL2-NOTCH-interaction} graphically, with the following
influences - $\bullet$ NOTCH family w.r.t ASCL2 with ASCL2 $->$ NOTCH-1/4. \par
\begin{table}[h!]
\begin{center}
\resizebox{\columnwidth}{!}{
\begin{tabular}{@{}ll@{}}
\multicolumn{2}{c}{\textsc{Unexplored combinatorial hypotheses}}\\
\hline
NOTCH family w.r.t ASCL2 & \\
NOTCH-1/4 & ASCL2\\
\hline
\end{tabular}
} % end of resizebox
\end{center}
\caption{$2^{nd}$ order combinatorial hypotheses between ASCL2
  and NOTCH family members.}
\label{tab:ASCL2-NOTCH-hypotheses}
\end{table}
%
%\clearpage
%
\subsubsection{ASCL2 - SLC}
\citet{Wang:2023molecular} indicate that syncytin-1 interacts with the SLC1A5 receptor on cells. \citet{Varberg:2021ascl2} found that knockdown of ASCL2 down regulated syncytin-1 receptor SLC1A5. In colorectal cancer cells treated with ETC-1922159, SLC family and ASCL2, were found to be down regulated and recorded independently. I was able to rank 2$^{nd}$ order combination of SLC family and ASCL2, that were down regulated. \par
Table \ref{tab:ASCL2-SLC-interaction} shows rankings of these
combinations. Followed by this is the unexplored
combinatorial hypotheses in table \ref{tab:ASCL2-SLC-hypotheses}
generated from analysis of the ranks in table
\ref{tab:ASCL2-SLC-interaction}. The table
\ref{tab:ASCL2-SLC-interaction} shows rankings of SLC family w.r.t
ASCL2. SLC39A8 - ASCL2 shows low ranking of 59 (laplace), 357 (linear) and 54 (rbf). SLC43A1 - ASCL2 shows low ranking of 62 (laplace), 99 (linear) and 117 (rbf). SLC25A27 - ASCL2 shows low ranking of 67 (laplace), 48 (linear) and 74 (rbf). SLC12A2 - ASCL2 shows low ranking of 91 (laplace), 451 (linear) and 318 (rbf). SLC7A2 - ASCL2 shows low ranking of 218 (laplace), 367 (linear) and 209 (rbf). SLC19A3 - ASCL2 shows low ranking of 292 (laplace), 136 (linear) and 87 (rbf). SLC19A1 - ASCL2 shows low ranking of 317 (laplace), 139 (linear) and 210 (rbf). SLC43A3 - ASCL2 shows low ranking of 520 (laplace), 150 (linear) and 382 (rbf). SLC1A4 - ASCL2 shows low ranking of 569 (laplace), 603 (linear) and 1301 (rbf). SLC25A26 - ASCL2 shows low ranking of 601 (laplace), 1035 (linear) and 745 (rbf). SLC4A7 - ASCL2 shows low ranking of 864 (laplace), 1185 (linear) and 915 (rbf). SLC28A3 - ASCL2 shows low ranking of 904 (laplace), 739 (linear) and 366 (rbf). SLC39A10 - ASCL2 shows low ranking of 1208 (laplace), 458 (linear) and 909 (rbf). SLC25A35 - ASCL2 shows low ranking of 1309 (laplace) and 1009 (rbf). SLC28A2 - ASCL2 shows low ranking of 1363 (laplace), 625 (linear) and 876 (rbf). SLC2A11 - ASCL2 shows low ranking of 1442 (laplace), 1343 (linear) and 1351 (rbf). SLC25A38 - ASCL2 shows low ranking of 1493 (laplace) and 1068 (linear). SLC35G1 - ASCL2 shows low ranking of 1522 (laplace) and 1506 (linear). SLC17A9 - ASCL2 shows low ranking of 1528 (laplace) and 1414 (linear). SLC7A8 - ASCL2 shows low ranking of 1268 (linear) and 1001 (rbf). SLC41A1 - ASCL2 shows low ranking of 1152 (linear) and 1016 (rbf). These rankings point to the synergy existing between the two components, which have been down regulated after the drug treatment. \par
Further, SLC25A15, SLC25A14, SLC18B1, SLC5A6, SLC12A8, SLC25A40, SLC35F2, SLC25A19, SLC26A2, SLC35E3, SLC25A32, SLC38A5 and SLC6A6 showed high ranking and might not be synergistically working with ASCL2, before treatment. \par
\begin{table}[ht!]
\begin{center}
\resizebox{\columnwidth}{!}{
\begin{tabular}{@{}lclc|lclc@{}}
\multicolumn{8}{c}{\textsc{Ranking SLC family VS ASCL2}}\\
\hline
\multicolumn{8}{c}{\textsc{Ranking of SLC family w.r.t ASCL2}} \\
                 & laplace  & linear & rbf &  & laplace  & linear & rbf \\
\hline
SLC39A8 - ASCL2 & 59 & 357 & 54 & SLC43A1 - ASCL2 & 62 & 99 & 117 \\
SLC25A27 - ASCL2 & 67 & 48 & 74 & SLC12A2 - ASCL2 & 91 & 451 & 318 \\
SLC7A2 - ASCL2 & 218 & 367 & 209 & SLC19A3 - ASCL2 & 292 & 136 & 87 \\
SLC19A1 - ASCL2 & 317 & 139 & 210 & SLC43A3 - ASCL2 & 520 & 150 & 382 \\
SLC1A4 - ASCL2 & 569 & 603 & 1301 & SLC25A26 - ASCL2 & 601 & 1035 & 745 \\
SLC4A7 - ASCL2 & 864 & 1185 & 915 & SLC28A3 - ASCL2 & 904 & 739 & 366 \\
SLC25A15 - ASCL2 & 1126 & 1621 & 1664 & SLC39A10 - ASCL2 & 1208 & 458 & 909 \\
SLC25A35 - ASCL2 & 1309 & 1703 & 1009 & SLC28A2 - ASCL2 & 1363 & 625 & 876 \\
SLC2A11 - ASCL2 & 1442 & 1343 & 1351 & SLC25A38 - ASCL2 & 1493 & 1068 & 2084 \\
SLC35G1 - ASCL2 & 1522 & 1506 & 1830 & SLC17A9 - ASCL2 & 1528 & 1414 & 1753 \\
SLC7A8 - ASCL2 & 1618 & 1268 & 1001 & SLC25A14 - ASCL2 & 1726 & 2341 & 2174 \\
SLC41A1 - ASCL2 & 1753 & 1152 & 1016 & SLC18B1 - ASCL2 & 1762 & 1461 & 2062 \\
SLC5A6 - ASCL2 & 1836 & 1606 & 2281 & SLC12A8 - ASCL2 & 1860 & 1018 & 1799 \\
SLC25A40 - ASCL2 & 1887 & 1987 & 1756 & SLC35F2 - ASCL2 & 1894 & 1585 & 2288 \\
SLC25A19 - ASCL2 & 1938 & 2108 & 1152 & SLC26A2 - ASCL2 & 2148 & 1756 & 2496 \\
SLC35E3 - ASCL2 & 2250 & 2656 & 1882 & SLC25A32 - ASCL2 & 2358 & 1622 & 2552 \\
SLC38A5 - ASCL2 & 2406 & 2344 & 2650 & SLC6A6 - ASCL2 & 2505 & 2456 & 1815 \\
\hline
\end{tabular}
} % end of resizebox
\end{center}
\caption{$2^{nd}$ order interaction ranking between ASCL2 VS SLC family
members.}
\label{tab:ASCL2-SLC-interaction}
\end{table}
One can also interpret the results of the table
\ref{tab:ASCL2-SLC-interaction} graphically, with the following
influences - $\bullet$ SLC family w.r.t ASCL2 with ASCL2 $->$ SLC-39A8 / 43A1 / 25A27 / 12A2 / 7A2 / 19A3 / 19A1 / 43A3 / 1A4 / 25A26 / 4A7 / 28A3 / 39A10 / 25A35 / 28A2 / 2A11 / 25A38 / 35G1 / 17A9 / 7A8 / 41A1. \par
\begin{table}[h!]
\begin{center}
\resizebox{\columnwidth}{!}{
\begin{tabular}{@{}ll@{}}
\multicolumn{2}{c}{\textsc{Unexplored combinatorial hypotheses}}\\
\hline
SLC family w.r.t ASCL2 & \\
SLC-39A8/43A1/25A27/12A2/7A2/19A3/19A1/43A3/1A4/25A26 & ASCL2\\
SLC-4A7/28A3/39A10/25A35/28A2/2A11/25A38/35G1/17A9/7A8/41A1 & ASCL2\\
\hline
\end{tabular}
} % end of resizebox
\end{center}
\caption{$2^{nd}$ order combinatorial hypotheses between ASCL2
  and SLC family members.}
\label{tab:ASCL2-SLC-hypotheses}
\end{table}
%
%\clearpage
%
\subsubsection{ASCL2 - SOX}
\citet{Allan:2021non} demonstrated that HIF1A activated a unique set of genes in oligodendrocyte progenitor cells (OPCs) through interaction with the transcription factor OLIG2 (which target ASCL2 and DLX3). These in turn, suppressed the oligodendrocyte regulator SOX10. ChIP-seq for HA-tagged ASCL2 in OPCs revealed that ASCL2 binds to upstream enhancers of SOX10 and reduce expression of SOX10 compared with control OPCs. In colorectal cancer cells treated with ETC-1922159, SOX family and ASCL2, were found to be down regulated and recorded independently. I was able to rank 2$^{nd}$ order combination of SOX family and ASCL2, that were down regulated. \par
Table \ref{tab:ASCL2-SOX-interaction} shows rankings of these
combinations. Followed by this is the unexplored
combinatorial hypotheses in table \ref{tab:ASCL2-SOX-hypotheses}
generated from analysis of the ranks in table
\ref{tab:ASCL2-SOX-interaction}. The table
\ref{tab:ASCL2-SOX-interaction} shows rankings of SOX family w.r.t
ASCL2. SOX8 - ASCL2 shows low ranking of 49 (laplace), 29 (linear) and 38 (rbf). SOX12 - ASCL2 shows low ranking of 203 (laplace), 590 (linear) and 405 (rbf). SOX1 - ASCL2 shows low ranking of 254 (laplace), 310 (linear) and 241 (rbf). These rankings point to the synergy existing between the two components, which have been down regulated after the drug treatment. \par
\begin{table}[ht!]
\begin{center}
\resizebox{\columnwidth}{!}{
\begin{tabular}{@{}lclc@{}}
\multicolumn{4}{c}{\textsc{Ranking SOX family VS ASCL2}}\\
\hline
\multicolumn{4}{c}{\textsc{Ranking of SOX family w.r.t ASCL2}} \\
                 & laplace  & linear & rbf \\
\hline
SOX8 - ASCL2 & 49 & 29 & 38 \\
SOX12 - ASCL2 & 203 & 590 & 405 \\
SOX1 - ASCL2 & 254 & 310 & 241 \\
\hline
\end{tabular}
} % end of resizebox
\end{center}
\caption{$2^{nd}$ order interaction ranking between ASCL2 VS SOX family
members.}
\label{tab:ASCL2-SOX-interaction}
\end{table}
One can also interpret the results of the table
\ref{tab:ASCL2-SOX-interaction} graphically, with the following
influences - $\bullet$ SOX family w.r.t ASCL2 with ASCL2 $->$ SOX-8/12/1. \par
\begin{table}[h!]
\begin{center}
\resizebox{\columnwidth}{!}{
\begin{tabular}{@{}ll@{}}
\multicolumn{2}{c}{\textsc{Unexplored combinatorial hypotheses}}\\
\hline
SOX family w.r.t ASCL2 & \\
SOX-8/12/1 & ASCL2\\
\hline
\end{tabular}
} % end of resizebox
\end{center}
\caption{$2^{nd}$ order combinatorial hypotheses between ASCL2
  and SOX family members.}
\label{tab:ASCL2-SOX-hypotheses}
\end{table}
%
%\clearpage
%
\subsubsection{ASCL2 - SNHG}
In colorectal cancer, \citet{Christensen:2016snhg16} show that SNHG16 is regulated by the WNT pathway. SNHG16 was found to be positively correlated to the expression of the WNT targets like ASCL2. In colorectal cancer cells treated with ETC-1922159, SNHG family and ASCL2, were found to be down regulated and recorded independently. I was able to rank 2$^{nd}$ order combination of SNHG family and ASCL2, that were down regulated. \par
Table \ref{tab:ASCL2-SNHG-interaction} shows rankings of these
combinations. Followed by this is the unexplored
combinatorial hypotheses in table \ref{tab:ASCL2-SNHG-hypotheses}
generated from analysis of the ranks in table
\ref{tab:ASCL2-SNHG-interaction}. The table
\ref{tab:ASCL2-SNHG-interaction} shows rankings of SNHG family w.r.t
ASCL2. SNHG3 - ASCL2 shows low ranking of 593 (laplace), 1415 (linear) and 1203 (rbf). SNHG15 - ASCL2 shows low ranking of 915 (laplace) and 1491 (rbf). SNHG10 - ASCL2 shows low ranking of 933 (laplace), 478 (linear) and 717 (rbf). SNHG17 - ASCL2 shows low ranking of 1108 (laplace) and 1388 (linear). SNHG16 - ASCL2 shows low ranking of 1398 (laplace) and 1011 (rbf). These rankings point to the synergy existing between the two components, which have been down regulated after the drug treatment. \par
Further, SNHG6, SNHG1, SNHG18, SNHG5, SNHG8 and SNHG7 showed high ranking and might not be synergistically working with ASCL2, before treatment. \par
\begin{table}[ht!]
\begin{center}
\resizebox{\columnwidth}{!}{
\begin{tabular}{@{}lclc@{}}
\multicolumn{4}{c}{\textsc{Ranking SNHG family VS ASCL2}}\\
\hline
\multicolumn{4}{c}{\textsc{Ranking of SNHG family w.r.t ASCL2}} \\
                 & laplace  & linear & rbf \\
\hline
SNHG3 - ASCL2 & 593 & 1415 & 1203 \\
SNHG15 - ASCL2 & 915 & 1888 & 1491 \\
SNHG10 - ASCL2 & 933 & 478 & 717 \\
SNHG17 - ASCL2 & 1108 & 1388 & 2096 \\
SNHG16 - ASCL2 & 1398 & 1972 & 1011 \\
SNHG6 - ASCL2 & 1582 & 2234 & 1728 \\
SNHG1 - ASCL2 & 1700 & 1744 & 1413 \\
SNHG18 - ASCL2 & 1730 & 2623 & 2302 \\
SNHG5 - ASCL2 & 1769 & 1945 & 411 \\
SNHG8 - ASCL2 & 2312 & 2167 & 1489 \\
SNHG7 - ASCL2 & 2417 & 1034 & 2005 \\
\hline
\end{tabular}
} % end of resizebox
\end{center}
\caption{$2^{nd}$ order interaction ranking between ASCL2 VS SNHG family
members.}
\label{tab:ASCL2-SNHG-interaction}
\end{table}
One can also interpret the results of the table
\ref{tab:ASCL2-SNHG-interaction} graphically, with the following
influences - $\bullet$ SNHG family w.r.t ASCL2 with ASCL2 $->$ SNHG-3/15/10/17/16. \par
\begin{table}[h!]
\begin{center}
\resizebox{\columnwidth}{!}{
\begin{tabular}{@{}ll@{}}
\multicolumn{2}{c}{\textsc{Unexplored combinatorial hypotheses}}\\
\hline
SNHG family w.r.t ASCL2 & \\
SNHG-3/15/10/17/16 & ASCL2\\
\hline
\end{tabular}
} % end of resizebox
\end{center}
\caption{$2^{nd}$ order combinatorial hypotheses between ASCL2
  and SNHG family members.}
\label{tab:ASCL2-SNHG-hypotheses}
\end{table}
%
%\clearpage
%
\subsubsection{ASCL2 - KIAA}
In colon cancer, \citet{Birkenkamp:2011repression} showed by immunofluorescence microscopy and western blotting that there was decreased protein expression of $\beta$-catenin and the WNT related stem cell marker ASCL2, upon KIAA1199 knockdown with construct KIAA1199-sh3303. In colorectal cancer cells treated with ETC-1922159, KIAA family and ASCL2, were found to be down regulated and recorded independently. I was able to rank 2$^{nd}$ order combination of KIAA family and ASCL2, that were down regulated. \par
Table \ref{tab:ASCL2-KIAA-interaction} shows rankings of these
combinations. Followed by this is the unexplored
combinatorial hypotheses in table \ref{tab:ASCL2-KIAA-hypotheses}
generated from analysis of the ranks in table
\ref{tab:ASCL2-KIAA-interaction}. The table
\ref{tab:ASCL2-KIAA-interaction} shows rankings of KIAA family w.r.t
ASCL2. KIAA0101 - ASCL2 shows low ranking of 73 (laplace), 350 (linear) and 178 (rbf). KIAA1524 - ASCL2 shows low ranking of 227 (laplace), 208 (linear) and 229 (rbf). KIAA1324 - ASCL2 shows low ranking of 370 (laplace), 218 (linear) and 52 (rbf). KIAA1586 - ASCL2 shows low ranking of 1065 (laplace), 936 (linear) and 793 (rbf). KIAA0586 - ASCL2 shows low ranking of 1189 (laplace) and 1304 (rbf). KIAA0020 - ASCL2 shows low ranking of 1406 (linear) and 955 (rbf). These rankings point to the synergy existing between the two components, which have been down regulated after the drug treatment. \par
Further, KIAA1244, KIAA1430, KIAA1257, KIAA1731 and KIAA1143 showed high ranking and might not be synergistically working with ASCL2, before treatment. \par
\begin{table}[ht!]
\begin{center}
\resizebox{\columnwidth}{!}{
\begin{tabular}{@{}lclc@{}}
\multicolumn{4}{c}{\textsc{Ranking KIAA family VS ASCL2}}\\
\hline
\multicolumn{4}{c}{\textsc{Ranking of KIAA family w.r.t ASCL2}} \\
                 & laplace  & linear & rbf \\
\hline
KIAA0101 - ASCL2 & 73 & 350 & 178 \\
KIAA1524 - ASCL2 & 227 & 208 & 229 \\
KIAA1324 - ASCL2 & 370 & 218 & 52 \\
KIAA1586 - ASCL2 & 1065 & 936 & 793 \\
KIAA0586 - ASCL2 & 1189 & 1564 & 1304 \\
KIAA1244 - ASCL2 & 1409 & 2640 & 2396 \\
KIAA1430 - ASCL2 & 1727 & 2495 & 1372 \\
KIAA1257 - ASCL2 & 1947 & 918 & 1673 \\
KIAA1731 - ASCL2 & 2134 & 2263 & 1148 \\
KIAA0020 - ASCL2 & 2156 & 1406 & 955 \\
KIAA1143 - ASCL2 & 2567 & 1781 & 2705 \\
\hline
\end{tabular}
} % end of resizebox
\end{center}
\caption{$2^{nd}$ order interaction ranking between ASCL2 VS KIAA family
members.}
\label{tab:ASCL2-KIAA-interaction}
\end{table}
One can also interpret the results of the table
\ref{tab:ASCL2-KIAA-interaction} graphically, with the following
influences - $\bullet$ KIAA family w.r.t ASCL2 with ASCL2 $->$ KIAA-0101/1524/1324/1586/0586/0020. \par
\begin{table}[h!]
\begin{center}
\resizebox{\columnwidth}{!}{
\begin{tabular}{@{}ll@{}}
\multicolumn{2}{c}{\textsc{Unexplored combinatorial hypotheses}}\\
\hline
KIAA family w.r.t ASCL2 & \\
KIAA-0101/1524/1324/1586/0586/0020 & ASCL2\\
\hline
\end{tabular}
} % end of resizebox
\end{center}
\caption{$2^{nd}$ order combinatorial hypotheses between ASCL2
  and KIAA family members.}
\label{tab:ASCL2-KIAA-hypotheses}
\end{table}
%
%\clearpage
%
\subsubsection{ASCL2 - FBXO}
Through various in vivo and in vitro assays, \citet{Tan:2011scffbxo22} show that FBXO22 is the substrate recognition subunit of the SCF$^{FBXO22}$ complex that polyubiquitylates KDM4A. They show that changes in FBXO22 levels can affect KDM4A protein levels. These then lead to changes in histone marks and changes in transcriptional levels of KDM4A’s target gene, ASCL2. In colorectal cancer cells treated with ETC-1922159, FBXO family and ASCL2, were found to be down regulated and recorded independently. I was able to rank 2$^{nd}$ order combination of FBXO family and ASCL2, that were down regulated. \par
Table \ref{tab:ASCL2-FBXO-interaction} shows rankings of these
combinations. Followed by this is the unexplored
combinatorial hypotheses in table \ref{tab:ASCL2-FBXO-hypotheses}
generated from analysis of the ranks in table
\ref{tab:ASCL2-FBXO-interaction}. The table
\ref{tab:ASCL2-FBXO-interaction} shows rankings of FBXO family w.r.t
ASCL2. FBXO5 - ASCL2 show low rankings of 351 (laplace), 316 (linear) and 105 (rbf). FBXO4 - ASCL2 show low rankings of 1466 (laplace), 710 (linear) and 347 (rbf). These rankings point to the synergy existing between the two components, which have been down regulated after the drug treatment. Further, FBXO41 showed high ranking and might not be synergistically working with ASCL2, before treatment. \par %
\begin{table}[ht!]
\begin{center}
\resizebox{\columnwidth}{!}{
\begin{tabular}{@{}lclc@{}}
\multicolumn{4}{c}{\textsc{Ranking FBXO family VS ASCL2}}\\
\hline
\multicolumn{4}{c}{\textsc{Ranking of FBXO family w.r.t ASCL2}} \\
                 & laplace  & linear & rbf \\
\hline
FBXO5 - ASCL2 & 351 & 316 & 105 \\
FBXO4 - ASCL2 & 1466 & 710 & 347 \\
FBXO41 - ASCL2 & 1685 & 2323 & 1224 \\
\hline
\end{tabular}
} % end of resizebox
\end{center}
\caption{$2^{nd}$ order interaction ranking between ASCL2 VS FBXO family
members.}
\label{tab:ASCL2-FBXO-interaction}
\end{table}
One can also interpret the results of the table
\ref{tab:ASCL2-FBXO-interaction} graphically, with the following
influences - $\bullet$ FBXO family w.r.t ASCL2 with ASCL2 $->$ FBXO-0101/1524/1324/1586/0586/0020. \par
\begin{table}[h!]
\begin{center}
\resizebox{\columnwidth}{!}{
\begin{tabular}{@{}ll@{}}
\multicolumn{2}{c}{\textsc{Unexplored combinatorial hypotheses}}\\
\hline
FBXO family w.r.t ASCL2 & \\
FBXO-0101/1524/1324/1586/0586/0020 & ASCL2\\
\hline
\end{tabular}
} % end of resizebox
\end{center}
\caption{$2^{nd}$ order combinatorial hypotheses between ASCL2
  and FBXO family members.}
\label{tab:ASCL2-FBXO-hypotheses}
\end{table}
%
%\clearpage
%
\subsubsection{ASCL2 - FAM}
\citet{Liang:2019molecular}, via electrophoretic mobility shift assay, validated the regulatory role of ACSL1 and ASCL2 in the regulation of FAM13A. There are range of family with sequence similarity members. In colorectal cancer cells treated with ETC-1922159, FAM family and ASCL2, were found to be down regulated and recorded independently. I was able to rank 2$^{nd}$ order combination of FAM family and ASCL2, that were down regulated. \par
Table \ref{tab:ASCL2-FAM-interaction} shows rankings of these
combinations. Followed by this is the unexplored
combinatorial hypotheses in table \ref{tab:ASCL2-FAM-hypotheses}
generated from analysis of the ranks in table
\ref{tab:ASCL2-FAM-interaction}. The table
\ref{tab:ASCL2-FAM-interaction} shows rankings of FAM family w.r.t
ASCL2. FAM111B - ASCL2 show low rankings of 9 (laplace), 71 (linear) and 16 (rbf). FAM161A - ASCL2 show low rankings of 106 (laplace), 86 (linear) and 93 (rbf). FAM72D - ASCL2 show low rankings of 112 (laplace), 284 (linear) and 161 (rbf). FAM201A - ASCL2 show low rankings of 162 (laplace), 398 (linear) and 341 (rbf). FAM216A - ASCL2 show low rankings of 265 (laplace), 453 (linear) and 198 (rbf).
FAM169A - ASCL2 show low rankings of 379 (laplace), 164 (linear) and 650 (rbf). FAM83D - ASCL2 show low rankings of 407 (laplace), 422 (linear) and 175 (rbf). FAM86A - ASCL2 show low rankings of 563 (laplace) and 1048 (rbf). FAM72A - ASCL2 show low rankings of 622 (laplace), 174 (linear) and 129 (rbf). FAM120C - ASCL2 show low rankings of 626 (laplace), 283 (linear) and 289 (rbf). FAM86C2P - ASCL2 show low rankings of 627 (laplace), 1279 (linear) and 1298 (rbf). FAM81A - ASCL2 show low rankings of 710 (laplace), 716 (linear) and 1124 (rbf). FAM227A - ASCL2 show low rankings of 717 (laplace), 327 (linear) and 254 (rbf). FAM86B1 - ASCL2 show low rankings of 757 (laplace), 1020 (linear) and 1466 (rbf). FAM96AP2 - ASCL2 show low rankings of 780 (laplace), 814 (linear) and 272 (rbf). FAM96A - ASCL2 show low rankings of 805 (laplace) and 1359 (linear). FAM98B - ASCL2 show low rankings of 889 (laplace), 951 (linear) and 277 (rbf). FAM117B - ASCL2 show low rankings of 1007 (laplace), 1114 (linear) and 867 (rbf). FAM131B - ASCL2 show low rankings of 1224 (laplace), 1289 (linear) and 295 (rbf). FAM221A - ASCL2 show low rankings of 1393 (laplace), 1158 (linear) and 1675.
FAM86JP - ASCL2 show low rankings of 1249 (linear) and 884 (rbf). These rankings point to the synergy existing between the two components, which have been down regulated after the drug treatment. Further, FAM41 showed high ranking and might not be synergistically working with ASCL2, before treatment. \par
Further, FAM89A, FAM185A, FAM117A, FAM136A, FAM86C1, FAM173B, FAM149B1, FAM210B, FAM208B, FAM178A, FAM122B and FAM98A showed high ranking and might not be synergistically working with ASCL2, before treatment. \par
\begin{table}[ht!]
\begin{center}
\resizebox{\columnwidth}{!}{
\begin{tabular}{@{}lclc|lclc@{}}
\multicolumn{8}{c}{\textsc{Ranking FAM family VS ASCL2}}\\
\hline
\multicolumn{8}{c}{\textsc{Ranking of FAM family w.r.t ASCL2}} \\
                 & laplace  & linear & rbf & & laplace  & linear & rbf \\
\hline
FAM111B - ASCL2 & 9 & 71 & 16 & FAM161A - ASCL2 & 106 & 86 & 93 \\
FAM72D - ASCL2 & 112 & 284 & 161 & FAM201A - ASCL2 & 162 & 398 & 341 \\
FAM216A - ASCL2 & 265 & 453 & 198 & FAM169A - ASCL2 & 379 & 164 & 650 \\
FAM83D - ASCL2 & 407 & 422 & 175 & FAM86A - ASCL2 & 563 & 2194 & 1048 \\
FAM72A - ASCL2 & 622 & 174 & 129 & FAM120C - ASCL2 & 626 & 283 & 289 \\
FAM86C2P - ASCL2 & 627 & 1279 & 1298 & FAM81A - ASCL2 & 710 & 716 & 1124 \\
FAM227A - ASCL2 & 717 & 327 & 254 & FAM86B1 - ASCL2 & 757 & 1020 & 1466 \\
FAM96AP2 - ASCL2 & 780 & 814 & 272 & FAM96A - ASCL2 & 805 & 1359 & 1543 \\
FAM98B - ASCL2 & 889 & 951 & 277 & FAM117B - ASCL2 & 1007 & 2478 & 2553 \\
FAM168A - ASCL2 & 1188 & 1114 & 867 & FAM131B - ASCL2 & 1224 & 1289 & 295 \\
FAM221A - ASCL2 & 1393 & 1158 & 1675 & FAM89A - ASCL2 & 1669 & 2217 & 2076 \\
FAM185A - ASCL2 & 1680 & 2182 & 2285 & FAM117A - ASCL2 & 1706 & 1801 & 1367 \\
FAM136A - ASCL2 & 1789 & 1209 & 2362 & FAM86C1 - ASCL2 & 1966 & 2592 & 1775 \\
FAM173B - ASCL2 & 2177 & 802 & 2511 & FAM86JP - ASCL2 & 2209 & 1249 & 884 \\
FAM149B1 - ASCL2 & 2258 & 2198 & 2667 & FAM210B - ASCL2 & 2362 & 1063 & 1657 \\
FAM208B - ASCL2 & 2435 & 1926 & 2393 & FAM178A - ASCL2 & 2461 & 2208 & 2703 \\
FAM122B - ASCL2 & 2691 & 2174 & 2504 & FAM98A - ASCL2 & 2736 & 2741 & 2658 \\
\hline
\end{tabular}
} % end of resizebox
\end{center}
\caption{$2^{nd}$ order interaction ranking between ASCL2 VS FAM family
members.}
\label{tab:ASCL2-FAM-interaction}
\end{table}
One can also interpret the results of the table
\ref{tab:ASCL2-FAM-interaction} graphically, with the following
influences - $\bullet$ FAM family w.r.t ASCL2 with ASCL2 $->$ FAM-111B / 161A / 72D / 201A / 216A / 169A / 83D / 86A / 72A / 120C / 86C2P / 81A / 227A / 86B1 / 96AP2 / 96A / 98D / 117B / 131B / 221A / 86JP. \par
\begin{table}[h!]
\begin{center}
\resizebox{\columnwidth}{!}{
\begin{tabular}{@{}ll@{}}
\multicolumn{2}{c}{\textsc{Unexplored combinatorial hypotheses}}\\
\hline
FAM family w.r.t ASCL2 & \\
FAM-111B/161A/72D/201A/216A/169A/83D/86A/72A/120C/86C2P & ASCL2\\
FAM-81A/227A/86B1/96AP2/96A/98D/117B/131B/221A/86JP & ASCL2\\
\hline
\end{tabular}
} % end of resizebox
\end{center}
\caption{$2^{nd}$ order combinatorial hypotheses between ASCL2
  and FAM family members.}
\label{tab:ASCL2-FAM-hypotheses}
\end{table}
%
%\clearpage
%
\subsubsection{ASCL2 - BCL}
In colorectal cancer cell lines, \citet{Legge:2019bcl} show for the first time that BCL3 acts as a co-activator of $\beta$-catenin/TCF-mediated transcriptional activity. They demonstrate that targeting BCL3 expression reduced $\beta$-catenin/TCF-dependent transcription and the expression of LGR5 and ASCL2. In colorectal cancer cells treated with ETC-1922159, BCL family and ASCL2, were found to be down regulated and recorded independently. I was able to rank 2$^{nd}$ order combination of BCL family and ASCL2, that were down regulated. \par
Table \ref{tab:ASCL2-BCL-interaction} shows rankings of these
combinations. Followed by this is the unexplored
combinatorial hypotheses in table \ref{tab:ASCL2-BCL-hypotheses}
generated from analysis of the ranks in table
\ref{tab:ASCL2-BCL-interaction}. The table
\ref{tab:ASCL2-BCL-interaction} shows rankings of BCL family w.r.t
ASCL2. BCL6B - ASCL2 shows low ranking of 240 (laplace), 235 (linear) and 172 (rbf). BCL11A - ASCL2 shows low ranking of 918 (laplace), 126 (linear) and 1248 (rbf). BCL11B - ASCL2 shows low ranking of 1245 (laplace), 810 (linear) and 1175 (rbf). These rankings point to the synergy existing between the two components, which have been down regulated after the drug treatment. Further, BCL9, BCL2L12 and BCL7A showed high ranking and might not be synergistically working with ASCL2, before treatment. \par
\begin{table}[ht!]
\begin{center}
\resizebox{\columnwidth}{!}{
\begin{tabular}{@{}lclc@{}}
\multicolumn{4}{c}{\textsc{Ranking BCL family VS ASCL2}}\\
\hline
\multicolumn{4}{c}{\textsc{Ranking of BCL family w.r.t ASCL2}} \\
                 & laplace  & linear & rbf \\
\hline
BCL6B - ASCL2 & 240 & 235 & 172 \\
BCL11A - ASCL2 & 918 & 126 & 1248 \\
BCL11B - ASCL2 & 1245 & 810 & 1175 \\
BCL9 - ASCL2 & 1698 & 2296 & 2011 \\
BCL2L12 - ASCL2 & 2263 & 2071 & 2499 \\
BCL7A - ASCL2 & 2706 & 2444 & 2317 \\
\hline
\end{tabular}
} % end of resizebox
\end{center}
\caption{$2^{nd}$ order interaction ranking between ASCL2 VS BCL family
members.}
\label{tab:ASCL2-BCL-interaction}
\end{table}
One can also interpret the results of the table
\ref{tab:ASCL2-BCL-interaction} graphically, with the following
influences - $\bullet$ BCL family w.r.t ASCL2 with ASCL2 $->$ BCL-6B/11A/11B. \par
\begin{table}[h!]
\begin{center}
\resizebox{\columnwidth}{!}{
\begin{tabular}{@{}ll@{}}
\multicolumn{2}{c}{\textsc{Unexplored combinatorial hypotheses}}\\
\hline
BCL family w.r.t ASCL2 & \\
BCL-6B/11A/11B & ASCL2\\
\hline
\end{tabular}
} % end of resizebox
\end{center}
\caption{$2^{nd}$ order combinatorial hypotheses between ASCL2
  and BCL family members.}
\label{tab:ASCL2-BCL-hypotheses}
\end{table}
%
%\clearpage
%
\subsubsection{ASCL2 - ATG}
In adult diffuse gliomas, \citet{Wang:2022ascl2} show that ASCL2 transcriptionally regulates the expression of ATG9B to maintain stemness properties. The established ASCL2-ATG9B axis demonstrate that it is crucial for maintaining the stemness phenotype and tumor progression, thus revealing a potential autophagy inhibition strategy. In colorectal cancer cells treated with ETC-1922159, ATG family and ASCL2, were found to be down regulated and recorded independently. I was able to rank 2$^{nd}$ order combination of ATG family and ASCL2, that were down regulated. \par
Table \ref{tab:ASCL2-ATG-interaction} shows rankings of these
combinations. Followed by this is the unexplored
combinatorial hypotheses in table \ref{tab:ASCL2-ATG-hypotheses}
generated from analysis of the ranks in table
\ref{tab:ASCL2-ATG-interaction}. The table
\ref{tab:ASCL2-ATG-interaction} shows rankings of ATG family w.r.t
ASCL2. ATG4C - ASCL2 shows low ranking of 334 (laplace), 638 (linear) and 939 (rbf). ATG10 - ASCL2 shows low ranking of 1439 (laplace) and 1166 (linear). These rankings point to the synergy existing between the two components, which have been down regulated after the drug treatment. Further, ATG3 showed high ranking and might not be synergistically working with ASCL2, before treatment. \par
\begin{table}[ht!]
\begin{center}
\resizebox{\columnwidth}{!}{
\begin{tabular}{@{}lclc@{}}
\multicolumn{4}{c}{\textsc{Ranking ATG family VS ASCL2}}\\
\hline
\multicolumn{4}{c}{\textsc{Ranking of ATG family w.r.t ASCL2}} \\
                 & laplace  & linear & rbf \\
\hline
ATG4C - ASCL2 & 334 & 638 & 939 \\
ATG10 - ASCL2 & 1439 & 1166 & 2148 \\
ATG3 - ASCL2 & 2172 & 2526 & 2010 \\
\hline
\end{tabular}
} % end of resizebox
\end{center}
\caption{$2^{nd}$ order interaction ranking between ASCL2 VS ATG family
members.}
\label{tab:ASCL2-ATG-interaction}
\end{table}
One can also interpret the results of the table
\ref{tab:ASCL2-ATG-interaction} graphically, with the following
influences - $\bullet$ ATG family w.r.t ASCL2 with ASCL2 $->$ ATG-4C/10. \par
\begin{table}[h!]
\begin{center}
\resizebox{\columnwidth}{!}{
\begin{tabular}{@{}ll@{}}
\multicolumn{2}{c}{\textsc{Unexplored combinatorial hypotheses}}\\
\hline
ATG family w.r.t ASCL2 & \\
ATG-4C/10 & ASCL2\\
\hline
\end{tabular}
} % end of resizebox
\end{center}
\caption{$2^{nd}$ order combinatorial hypotheses between ASCL2
  and ATG family members.}
\label{tab:ASCL2-ATG-hypotheses}
\end{table}
%
%\clearpage
%
\subsubsection{ASCL2 - ARHGAP}
In breast cancer, \citet{Han:2023arhgap25} identified a feedback loop that downregulated ARHGAP25, which can promote activation of the WNT/$\beta$-catenin pathway. Because of this there is an increase in ASCL2 expression, which negatively regulates ARHGAP25 expression at transcriptional level and contributes to breast cancer progression. In colorectal cancer cells treated with ETC-1922159, ARHGAP family and ASCL2, were found to be down regulated and recorded independently. I was able to rank 2$^{nd}$ order combination of ARHGAP family and ASCL2, that were down regulated. \par
Table \ref{tab:ASCL2-ARHGAP-interaction} shows rankings of these
combinations. Followed by this is the unexplored
combinatorial hypotheses in table \ref{tab:ASCL2-ARHGAP-hypotheses}
generated from analysis of the ranks in table
\ref{tab:ASCL2-ARHGAP-interaction}. The table
\ref{tab:ASCL2-ARHGAP-interaction} shows rankings of ARHGAP family w.r.t
ASCL2. ARHGAP11A - ASCL2 shows low ranking of 31 (laplace), 10 (linear) and 15 (rbf). ARHGAP11B - ASCL2 shows low ranking of 306 (laplace), 640 (linear) and 270 (rbf). ARHGAP33 - ASCL2 shows low ranking of 1048 (laplace), 1007 (linear) and 535 (rbf). These rankings point to the synergy existing between the two components, which have been down regulated after the drug treatment. Further, ARHGAP19 showed high ranking and might not be synergistically working with ASCL2, before treatment. \par
\begin{table}[ht!]
\begin{center}
\resizebox{\columnwidth}{!}{
\begin{tabular}{@{}lclc@{}}
\multicolumn{4}{c}{\textsc{Ranking ARHGAP family VS ASCL2}}\\
\hline
\multicolumn{4}{c}{\textsc{Ranking of ARHGAP family w.r.t ASCL2}} \\
                 & laplace  & linear & rbf \\
\hline
ARHGAP11A - ASCL2 & 31 & 10 & 15 \\
ARHGAP11B - ASCL2 & 306 & 640 & 270 \\
ARHGAP33 - ASCL2 & 1048 & 1007 & 535 \\
ARHGAP19 - ASCL2 & 1871 & 2288 & 2176 \\
\hline
\end{tabular}
} % end of resizebox
\end{center}
\caption{$2^{nd}$ order interaction ranking between ASCL2 VS ARHGAP family
members.}
\label{tab:ASCL2-ARHGAP-interaction}
\end{table}
One can also interpret the results of the table
\ref{tab:ASCL2-ARHGAP-interaction} graphically, with the following
influences - $\bullet$ ARHGAP family w.r.t ASCL2 with ASCL2 $->$ ARHGAP-11A/11B/33. \par
\begin{table}[h!]
\begin{center}
\resizebox{\columnwidth}{!}{
\begin{tabular}{@{}ll@{}}
\multicolumn{2}{c}{\textsc{Unexplored combinatorial hypotheses}}\\
\hline
ARHGAP family w.r.t ASCL2 & \\
ARHGAP-11A/11B/33 & ASCL2\\
\hline
\end{tabular}
} % end of resizebox
\end{center}
\caption{$2^{nd}$ order combinatorial hypotheses between ASCL2
  and ARHGAP family members.}
\label{tab:ASCL2-ARHGAP-hypotheses}
\end{table}
%
%\clearpage
%
\section{Conclusion}
Presented here are a range of multiple synergistic ASCL2 $2^{nd}$ order combinations that were
ranked via a machine learning based search engine. Via majority voting across
the ranking methods, it was possible to find plausible unexplored
synergistic combinations of ASCL2-X that might be prevalent in CRC cells after
treatment with ETC-1922159 drug. \par

%
%%%END OF MAIN TEXT%%%
\section*{Conflict of interest}
There are no conflicts to declare.

\section*{Author's contributions}
    Concept, design, in silico implementation - SS. Analysis and interpretation
    of results - SS. Manuscript writing - SS. Manuscript revision -
    SS. Approval of manuscript - SS

\section*{Acknowledgements}
  Special thanks to Mrs. Rita Sinha and Mr. Prabhat
  Sinha for supporting the author financially, without which this work
  could not have been made possible.

\section*{Source of Data}
Data used in this research work was released in a publication in
\citet{Madan:2016}.

\section{References}

\bibliography{elsarticle-template-3-modified-ASCL2}
%\bibliography{manuscript_11_modified}
%\bibliography{manuscript_11}

\end{document}